\documentclass[a4paper]{spie}  
\usepackage[]{graphicx}
\usepackage{amssymb,amsmath,psfrag,url}
\usepackage{verbatim}

\newcommand{\Bolivarallee}{Boliva\hspace{-0.1mm}r\hspace{0.15mm}a\hspace{-0.1mm}llee}
\newcommand{\Takustrasse}{Taku\hspace{0.25mm}s\hspace{-0.1mm}tra{\ss}e}

\title{
Finite-element based electromagnetic field simulations:
Benchmark results for isolated structures
}

\author{
Sven~Burger,\supit{\,ab}
Lin~Zschiedrich,\supit{\,a}
Jan~Pomplun,\supit{\,a}
Frank~Schmidt,\supit{\,ab}
\skiplinehalf
\supit{a}
JCMwave GmbH,
\Bolivarallee~22, 
D\,--\,14\,050 Berlin,
Germany
\smallskip\\
\supit{b}
Zuse Institute Berlin\,(ZIB),
\Takustrasse~7,
D\,--\,14\,195 Berlin,
Germany
}
\authorinfo{
Corresponding author: S. Burger\\
URL: http://www.zib.de\\
URL: http://www.jcmwave.com
}

\begin{document}
\maketitle

\noindent
This paper will be published in Proc.~SPIE Vol.~{\bf 8880}
(2013) 88801Z, ({\it Photomask Technology 2013}, DOI: 10.1117/12.2026213), 
and is made available 
as an electronic preprint with permission of SPIE. 
One print or electronic copy may be made for personal use only. 
Systematic or multiple reproduction, distribution to multiple 
locations via electronic or other means, duplication of any 
material in this paper for a fee or for commercial purposes, 
or modification of the content of the paper are prohibited.
\begin{abstract}
We use a finite-element method to obtain highly converged results for a 
nano-optical light scattering setup with a non-periodic geometry. \vspace{1cm}
\end{abstract}


\section{Introduction}

Electromagnetic field (EMF) simulations are important for design of experimental setups and 
understanding of experimental  
results in science and technology. 
As prominent example, recently in nanoscience and -technology and in related semiconductor 
industries, efficient ({\it = precise \& fast}) EMF simulations are given significant  
attention~\cite{ITRS2012M}: 
Performance of lithography at deep ultraviolet (DUV)  and extreme ultraviolet (EUV) 
wavelengths can be pushed using computational methods~\cite{Lai2012aot}. 
Further, computational methods are an integral part of optical metrology setups in this field~\cite{Pang2012aot}. 
EMF simulation of {\it non-periodic} patterns at high numerical resolution is gaining importance 
for metrology applications like in-die photomask registration and metrology and critical-dimension (CD) metrology
and for many lithography applications handling finite patterns or isolated assist features.   

Several numerical methods are used for EMF simulations, including, e.g.,
finite-difference time-domain methods (FDTD), rigorous coupled wave analysis (RCWA) 
and finite-element methods (FEM)~\cite{Burger2005bacus}. 
This paper concentrates on FEM in an implementation by JCMwave~\cite{Pomplun2007pssb}.  
In previous contributions these finite-element methods have been applied to various scientific and 
technological fields~\cite{Burger2013pw}. 
These include also computational lithography tasks 
like investigations of 3D effects, source-mask optimization, analysis of the impact of 
line edge/width roughness and defects, as well as to computational metrology tasks like 
cricital dimension (CD) metrology at EUV wavelengths and metrology of 3D 
patterns~\cite{Burger2007om,PoBuSc08b,Pomplun2010bacus,Burger2011bacus,Kleemann2011eom3}.

In order to demonstrate FEM performance for 
simulation of non-periodic patterns, 
in this contribution we revisit a specific benchmark example for electromagnetic field solvers.
The benchmark setup consists of computing the near field in an isolated (i.e., non-periodic) pattern
illuminated by a plane wave. 
We demonstrate numerical convergence of our method with various numerical parameters, in 
quantitative agreement with results from the literature~\cite{Lalanne2007jeos}, and we 
demonstrate an advance in performance by several orders of magnitude when compared to results 
from the literature~\cite{Lalanne2007jeos}. 
Previous benchmarks including this solver mainly concentrated on periodic patterns 
and other devices~\cite{Burger2005bacus,Burger2007bacus,Burger2008bacus,Hoffmann2009spie,Maes2013oe}.

This paper is structured as follows: 
The benchmark setup is described in Section~\ref{section_problem_description}, 
the numerical method and obtained results are presented in Section~\ref{section_numerical}. 
Selected numerical values obtained in the convergence studies are tabulated in the Appendix. 

\section{Benchmark: Non-periodic diffraction problem}
\label{section_problem_description}

We revisit a numerical 
benchmark problem for an isolated (non-periodic) setup  which has been described by Lalanne {\it et al.}~\cite{Lalanne2007jeos}:
The problem models an isolated, sub-wavelength slit in a silver film on a substrate with a neighboring, parallel 
groove in the silver film. 
This setup is illuminated by a plane wave at perpendicular incidence from above and with in-plane electric field
polarization (resp.~out-of-plane magnetic field polarization).
The energy flux of light, $S$, transmitted 
through the slit to a detector region of width $w_\mathrm{d}$, placed a distance $h_\mathrm{d}$
below the slit is detected and normalized 
to the energy flux $S_0$ through the slit, computed in a second simulation where the groove is not present.

\begin{figure}
\begin{center}
  \psfrag{d}{\sffamily $d$}
  \psfrag{wl}{\sffamily $w_\mathrm{x}$}
  \psfrag{wr}{\sffamily $w_\mathrm{x}$}
  \psfrag{wg}{\sffamily $w_\mathrm{g}$}
  \psfrag{ws}{\sffamily $w_\mathrm{s}$}
  \psfrag{wd}{\sffamily $w_\mathrm{d}$}
  \psfrag{hg}{\sffamily $h_\mathrm{g}$}
  \psfrag{hs}{\sffamily $h_\mathrm{s}$}
  \psfrag{hsu}{\sffamily $h_\mathrm{a}$}
  \psfrag{hsb}{\sffamily $h_\mathrm{d}$}
  \psfrag{air}{\sffamily air}
  \psfrag{silver}{\sffamily silver}
  \psfrag{substrate}{\sffamily substrate}
  \includegraphics[width=.6\textwidth]{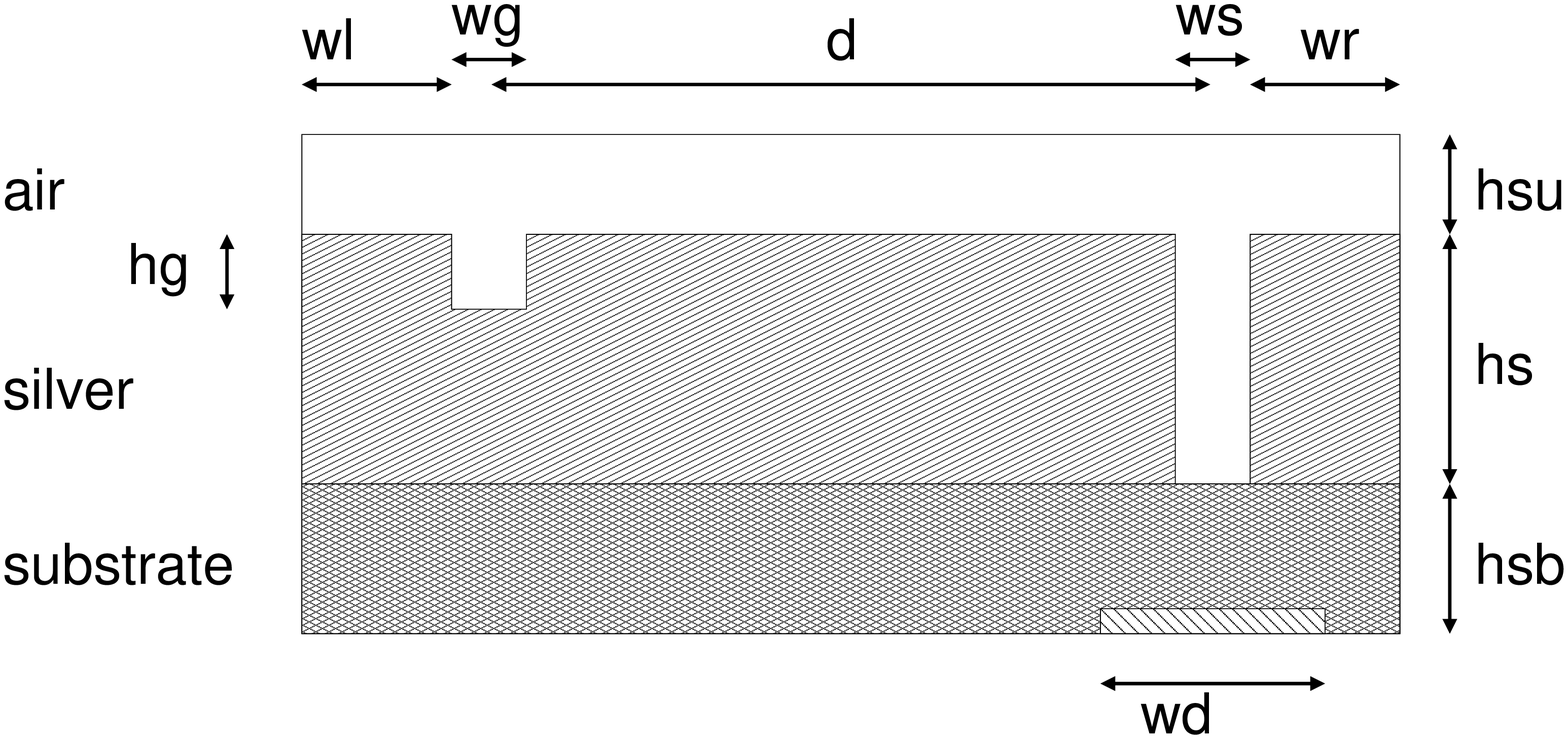}
  \includegraphics[width=.38\textwidth]{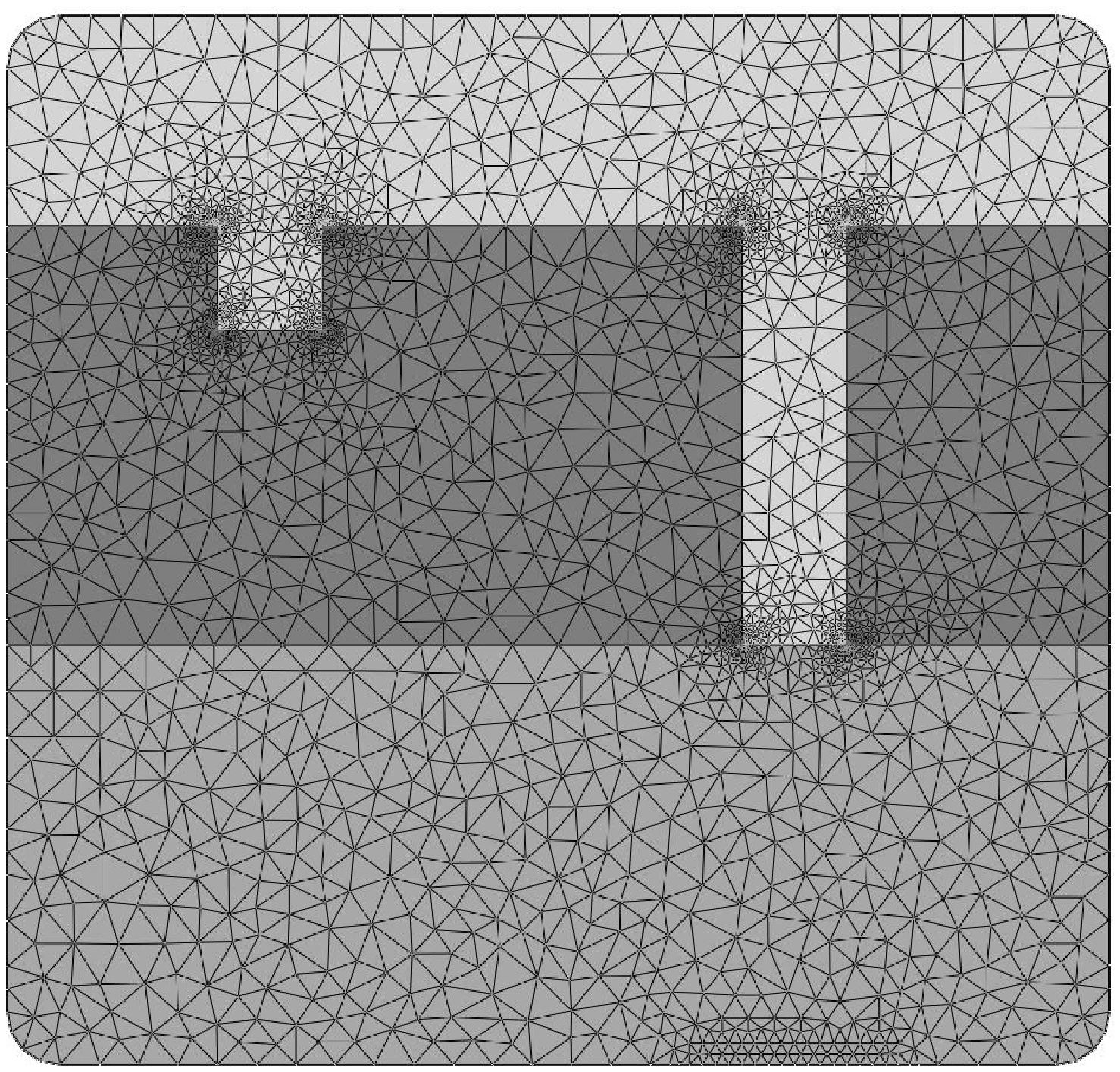}
  \caption{{\it Left:} Schematics of the geometry of a slit and a groove in a silver layer on a substrate.  
           At the left and right boundary of the depicted geometry, 
           the layer stack of substrate, silver film and air is extended to infinity. 
           The physical parameter values (geometrical parameters 
           $w_\mathrm{g}$, $w_\mathrm{s}$, $w_\mathrm{d}$, $h_\mathrm{g}$, $h_\mathrm{s}$, $h_\mathrm{d}$
           material and source parameters) are given in Table~\ref{table_parameters}. 
           The distances of the computational domain boundary from the scattering slit and groove, 
         $w_\mathrm{x}$ and $h_\mathrm{a}$, are numerical parameters which are varied in the numerical tests, 
         cf.~Table~\ref{table_my_results}. 
           {\it Right:} Visualization of a mesh discretizing the geometry of the slit-groove setup. 
            Note the finer discretization at the corners of the slit and of the groove cross-sections. 
            }
\label{fig_schematics}
\end{center}
\end{figure}

\begin{table}[b]
\begin{center}
\begin{tabular}{|l|l||l|l|}
\hline
$d$ & 500\,nm & $\lambda_0$ & {852\,nm}\\
$w_\mathrm{s} = w_\mathrm{g}$ & 100\,nm & &\\
$w_\mathrm{d}$ & 200\,nm & $\varepsilon_\mathrm{Ag}$ & -33.22 + 1.170i\\
$h_\mathrm{g}$ & 100\,nm & $\varepsilon_\mathrm{substrate}$ & 2.25\\
$h_\mathrm{s} = h_\mathrm{d}$ & 400\,nm & $\varepsilon_\mathrm{air}$ & 1\\
\hline
\end{tabular}
\caption{Physical parameter setting 
for the benchmark simulations. 
Dimensional parameters (compare Fig.~\ref{fig_schematics}):
slit-groove distance, $d$, slit and groove widths, $w_\mathrm{s}$ and $w_\mathrm{g}$,
detector width,  $w_\mathrm{d}$, 
slit and groove height, $h_\mathrm{s}$, $h_\mathrm{g}$, 
and detector distance from slit, $h_\mathrm{d}$.
Source vacuum wavelength $\lambda_0$ and 
complex material relative permittivities $\varepsilon$.
}
\label{table_parameters}
\end{center}
\end{table}

Due to the geometrical, source and material properties 
plasmonic effects lead to a very critical dependence of normalized transmission $S/S_0$ 
on the physical parameters. 
In essence, the source field is scattered at the groove and slit structures. This excites highly peaked surface fields 
which propagate along the surface, interfere and are re-scattered at the slit and groove structures. 
The relatively simple physical setup combined with the critical electromagnetic behavior makes this configuration 
a very interesting benchmark case. 
The choice of the geometry was originally also motivated by experimental results~\cite{Lalanne2007jeos,gay2006optical}

The geometry of the slit-groove configuration is depicted schematically in Fig.~\ref{fig_schematics}. 
The corresponding parameters, and the material and source properties are given in Table~\ref{table_parameters}.

In the original benchmark publication~\cite{Lalanne2007jeos} different numerical methods and implementations 
are used to obtain twelve different numerical results. 
The average result is $S/S_0\approx 2.18$ with a standard deviation of about 0.03 ({\it cf.} Fig.~5 in~\cite{Lalanne2007jeos}).
Internal convergence (i.e., convergence toward a specific {\it quasi-exact} result for each method) 
is observed with all methods. However, different levels of accuracy are reached, best internal relative accuracies are up to about 
 $10^{-5}$ which is significantly lower than the deviations of the best converged results of the different methods from each other. 
Table~\ref{table_results_lalanne} summarizes some of these results.

\section{Numerical method and results}
\label{section_numerical}
\subsection{Finite element method}
For rigorous simulations of the electromagnetic near field we use the 
FEM Maxwell solver JCMsuite.
This solver includes implementations of higher-order edge-elements, self-adaptive meshing, 
and fast solution algorithms for solving time-harmonic Maxwell's 
equations~\cite{Zschiedrich03,zschiedrich2007goaloriented,Zschiedrich2008al,Pomplun2007pssb}. 

\begin{figure}[b]
\begin{center}
  \psfrag{Rel. error}{\sffamily $\Delta(S/S_0)_{\mathrm{rel}}$ }
  \psfrag{N}{\sffamily N}
  \psfrag{N [1e5]}{\sffamily N [$10^5$]}
  \includegraphics[width=.47\textwidth]{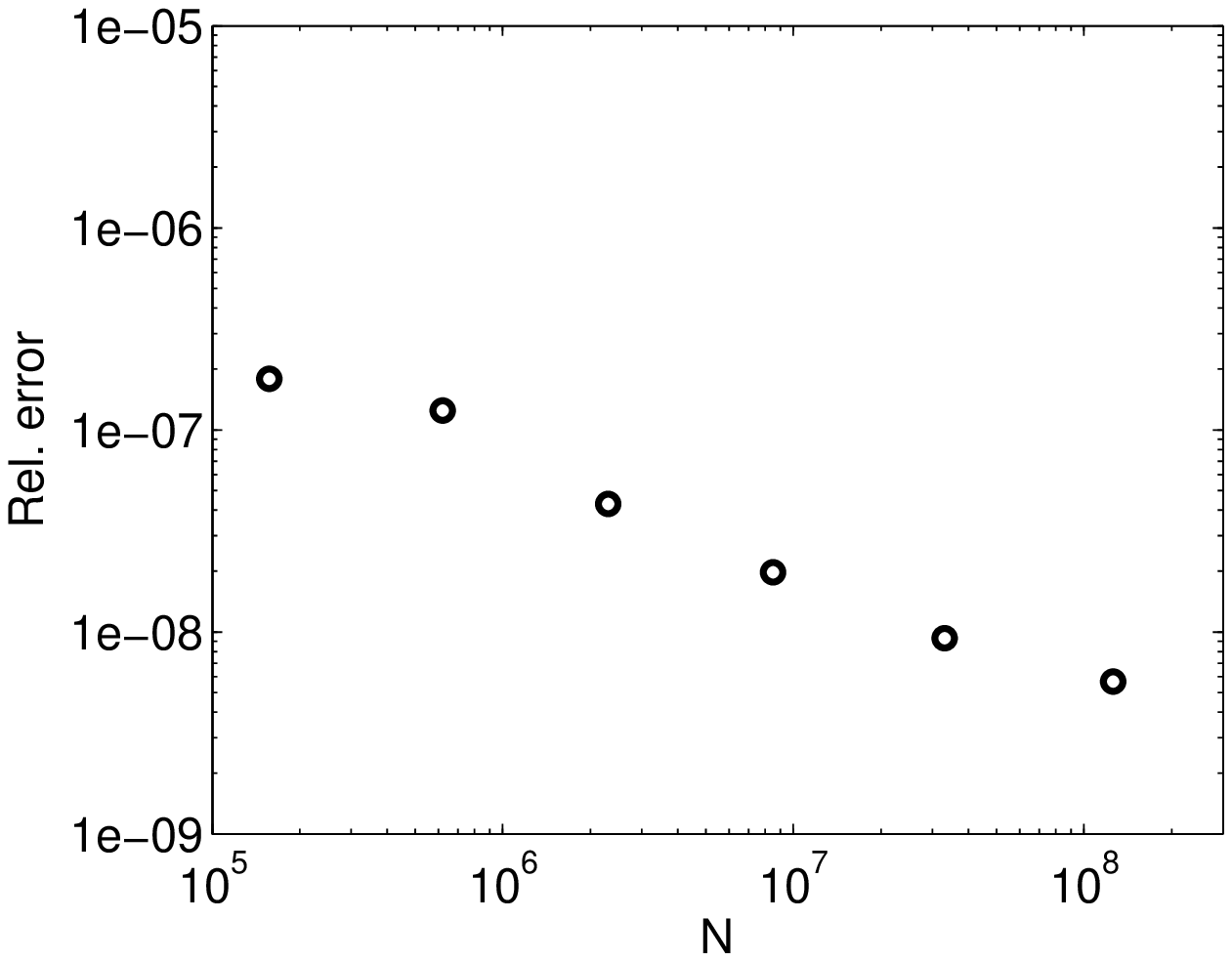}
  \hfill
  \includegraphics[width=.47\textwidth]{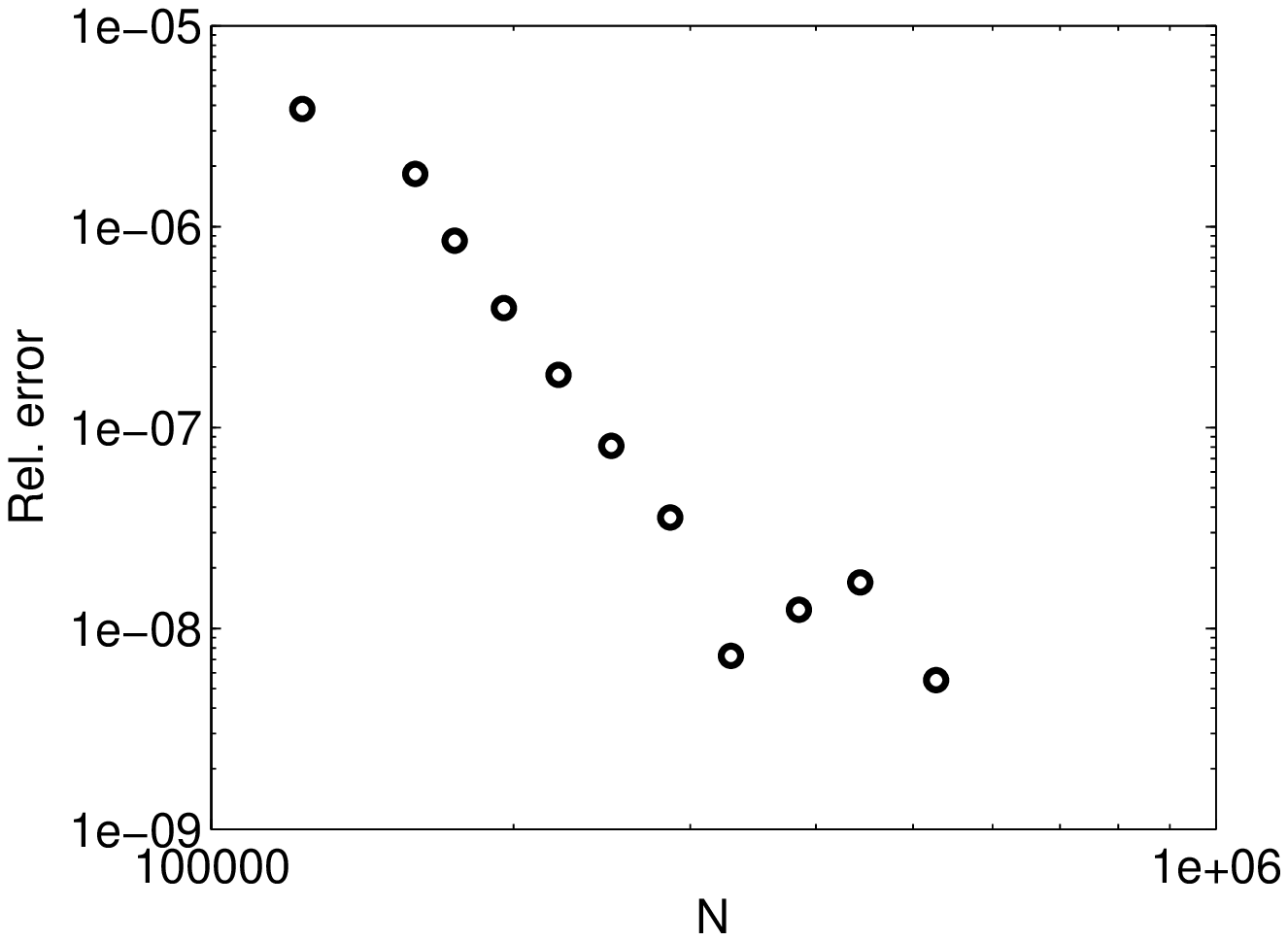}
  \caption{Convergence: Dependence of the relative numerical error, $\Delta(S/S_0)_{\mathrm{rel}}$,
  on the number of unknowns in the FEM problem, $N$. The data points are 
generated using regular mesh refinement ({\it left}) and an automatic, adaptive 
mesh refinement strategy ({\it right}). 
Finite elements of polynomial order $p=4$ have been used, compare also Table~\ref{table_my_results}.
}
\label{fig_convergence_refinement}
\end{center}
\end{figure}

Briefly, the simulations are performed as follows:
a scripting language (Matlab) automatically iterates the numerical input parameters
for fixed physical project parameters. 
The build-in mesh generator triangulates the geometries with and without the groove. 
Figure~\ref{fig_schematics} (right) shows a graphical representation of a mesh. 
The FEM solver computes the near fields, and in a post-process computes from these the integrals over the electromagnetic 
field energy flux densities (Poynting vector field) in the detection region, $S$ and $S_0$. 

Numerical parameters in the simulation setup are the polynomial degree of the finite-element ansatz functions, $p$, 
the number of (adaptive or global) grid refinement steps, $n_\mathrm{adapt}$, resp.~$n_\mathrm{reg}$, 
the offset of the structures in the computational 
domain from the boundaries of the computational domain,  $w_\mathrm{x}$ (with $h_\mathrm{a} = w_\mathrm{x}$), 
a minimum mesh size for local prerefinement of the 
initial meshes around corners of the  geometry, $dx$, 
and a sidelength-constraint, $SLC$, defining the maximum sidelength of the triangles of the initial 
mesh (prior to the  $n_\mathrm{adapt}+n_\mathrm{reg}$ mesh refinement steps).
Discretizing Maxwell's equations for given numerical parameters yields a linear system of equations with $N$ unknowns. 
Typically, when only one numerical parameter is varied, an increase of $N$ yields an increase of accuracy of the 
discrete solution, at least when a convergent regime is reached. 
In our case we have several parameters at hand, $p$, $n_\mathrm{adapt}$, $n_\mathrm{reg}$,
$w_\mathrm{x}$, $dx$, $SLC$.  All of these influence $N$, but influence of these parameters on the accuracy of the 
discrete solution should be quite different:
Increase of FEM degree $p$ is expected to greatly improve accuracy at significantly increased numerical 
costs, $N$. 
For field distributions with highly localized field peaks (as is present at the given metal 
edges and corners) accuracy improvement through local mesh refinement, influenced by parameters $n_\mathrm{adapt}$ and 
$dx$ is expected to be dominant at relatively low impact on $N$, compared to regular mesh refinement, 
controlled by parameters $n_\mathrm{reg}$ and $SLC$. 
For the computational domain offset, no influence on the accuracy is expected as long as the boundary conditions
are implemented as reflection-free boundaries ({\it transparent} boundaries, 
relatized by so-called {\it perfectly matched layers}, PML~\cite{Zschiedrich03}). 

For each set of numerical parameters, the result $S/S_0$ is saved, together with the number of unknowns $N$ and the 
total CPU time for computing $S$ and $S_0$. Computations are performed on a standard computer with extended RAM (necessary 
for computations with high $N$). 

\subsection{Numerical results}

FEM computations using {\it JCMsuite} yield a result of $S/S_0 \approx 2.19882594$. 
In order to estimate the numerical error of our computed result and in order to show that the result does not 
depend on possible systematic errors introduced by the method we have performed computations for a variety of 
numerical parameter sets. 
Some of the results are listed in the Appendix, in Table~\ref{table_my_results}.
As no reliable reference value at sufficient accuracy is at hand we use 
the result of $S/S_0$ for the numerical parameter data set with highest finite element degree $p=9$ 
and rather fine meshing 
as {\it quasi-exact} result, $S/S_{0,\mathrm{qe}}$. 
The relative error of a numerical result for $S/S_0$ is then defined as 
$\Delta(S/S_0)_{\mathrm{rel}} = |S/S_0-S/S_{0,\mathrm{qe}}|/(S/S_{0,\mathrm{qe}})$.

{\bf Convergence with mesh refinement} 
Figure~\ref{fig_convergence_refinement} shows how the relative error decreases with increase in computational effort, reached 
by regular mesh refinement (Fig.~\ref{fig_convergence_refinement}, left) and adaptive 
 mesh refinement (Fig.~\ref{fig_convergence_refinement}, right). 
For adaptive mesh refinement, a local {\it error-estimator} checks locally smoothness of the solution and 
yields local mesh refinement for only those elements with relatively high estimated errors. 
For regular mesh refinement, all triangles are refined to four smaller triangles, each. Both methods yield 
highly accurate results, however for adaptive mesh refinement computational costs are greatly reduced when compared 
to global mesh refinement. 
The polynomial degree of the finite element ansatz functions (FEM degree) was fixed to $p=4$ in these simulations (compare 
the first two data sets in Table~\ref{table_my_results}).

\begin{figure}[b]
\begin{center}
  \psfrag{Rel. error}{\sffamily $\Delta(S/S_0)_{\mathrm{rel}}$ }
  \psfrag{p}{\sffamily $p$}
  \psfrag{dx [nm]}{\sffamily $dx$ [nm]}
  \includegraphics[width=.47\textwidth]{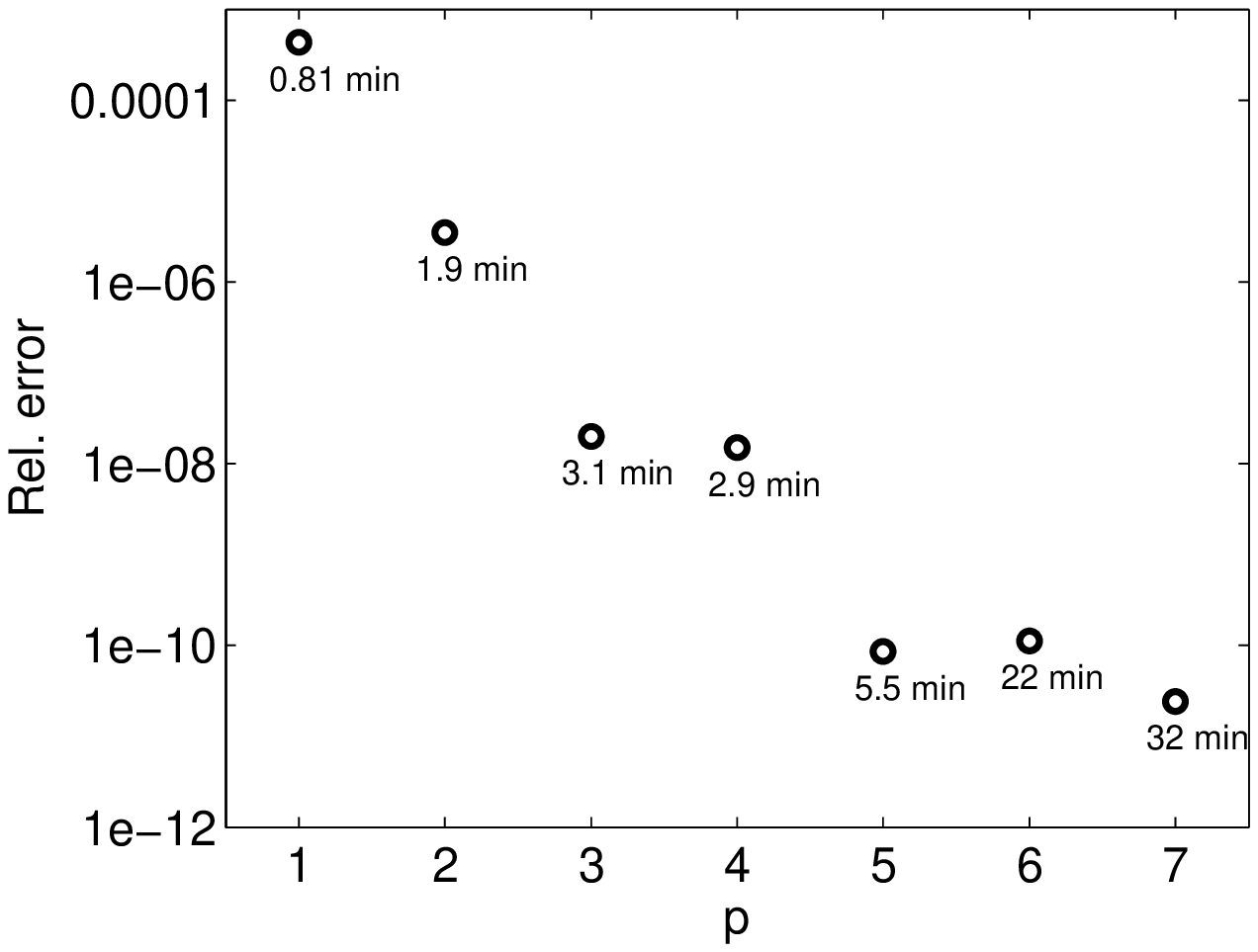}
  \hfill
  \includegraphics[width=.47\textwidth]{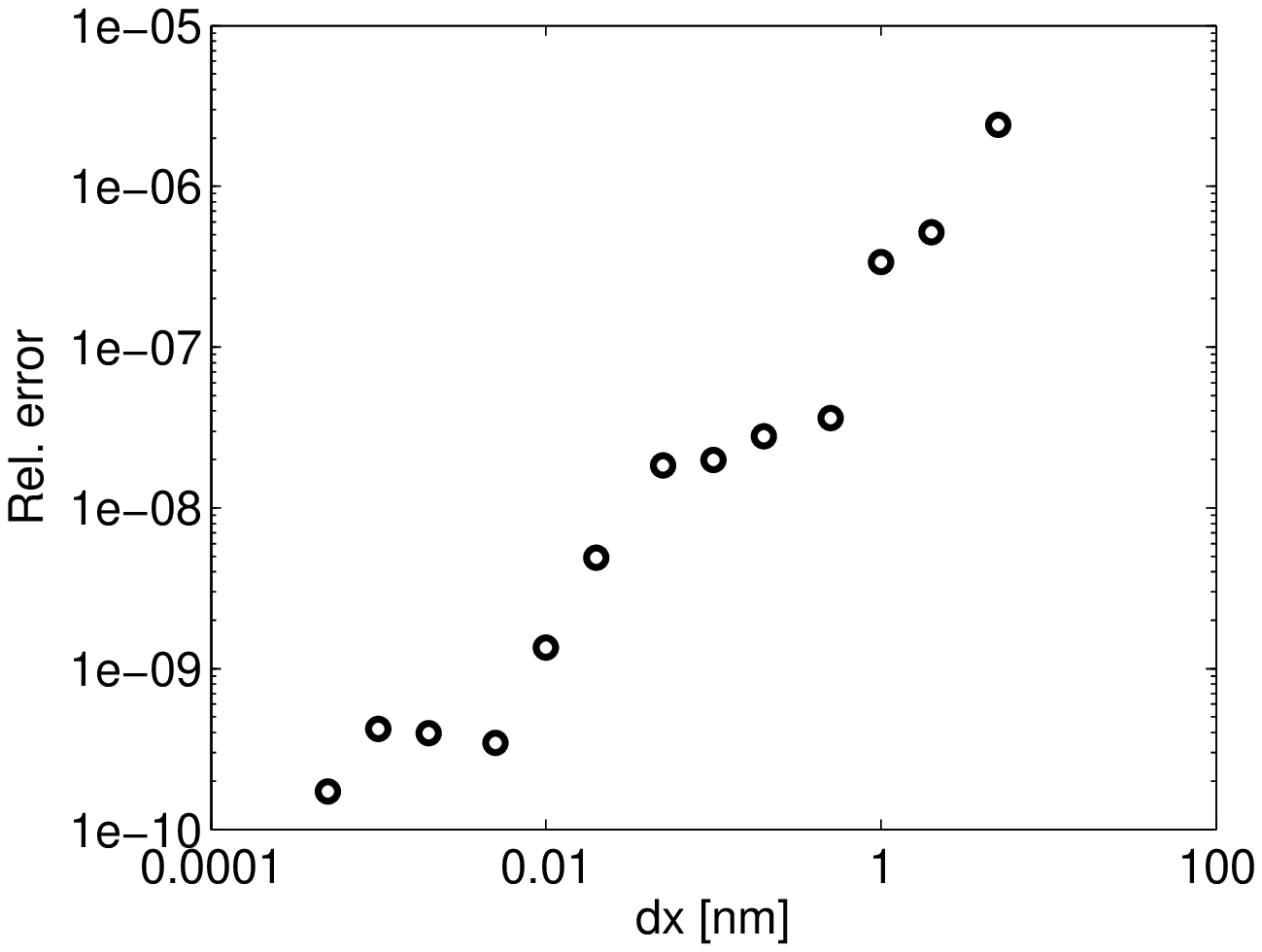}
  \caption{Convergence: {\it Left:} Dependence of the relative numerical error, $\Delta(S/S_0)_{\mathrm{rel}}$,
  on the polynomial order of the finite element ansatz functions, $p$ for  $n_\mathrm{adapt} = 1$ and $n_\mathrm{reg}=2$
mesh refinement steps. 
Total computation times on a standard PC are indicated. 
{\it Right:} 
Dependence of the relative numerical error, $\Delta(S/S_0)_{\mathrm{rel}}$, on the minimum mesh size at 
around the corners of the edges of the slit and of the groove, $dx$.  
Please, compare also Table~\ref{table_my_results}.
}
\label{fig_convergence_p_dx}
\end{center}
\end{figure}

{\bf Convergence with FEM degree} 
Convergence of numerical error with FEM degree is displayed in Figure~\ref{fig_convergence_p_dx} (left). 
For these simulations the mesh refinement strategy was fixed to always the same initial mesh and three 
adaptive refinement steps. 
FEM degree $p$ is varied between $p=1$ and $p=7$. 
For $p>4$ a relative error below $10^{-9}$ is reached, $\Delta(S/S_0)_{\mathrm{rel}} < 10^{-9}$. 

{\bf Convergence with local prerefinement} 
Figure~\ref{fig_convergence_p_dx} (right) shows convergence of the numerical error when both, $p$ and mesh refinement strategy 
are fixed, but the initial mesh is locally prerefined around the slit and groove edges with local mesh 
fineness $dx$.
Numerical accuracy shows a strong dependence on this parameter.
Relative numerical errors below  $10^{-9}$ are reached for $dx<0.01\,$nm (using FEM degree $p=7$, 
three regular mesh refinement steps and two adaptive mesh refinement steps). 
Very good local accuracy at critical positions in the near field obviously has a great impact on 
the accuracy of energy flux obtained at the detector position. 
This observation is similar to the observation of the advantage of adaptive mesh refinement versus 
global mesh refinement. 
It might also explain the relatively large error ranges and absolute deviations of the different methods 
highlighted in the literature~\cite{Lalanne2007jeos}, since here mainly non-adaptive methods have been used 
or adaptivity was restricted to relatively coarse sampling
(in the case of Fourier modal methods with complex coordinate transforms, MM3~\cite{Lalanne2007jeos}). 

\begin{figure}[b]
\begin{center}
  \psfrag{Rel. error}{\sffamily $\Delta(S/S_0)_{\mathrm{rel}}$ }
  \psfrag{SLC [nm]}{\sffamily $SLC$ [nm]}
  \psfrag{Offset [nm]}{\sffamily offset $w_\mathrm{x}$ [nm]}
  \includegraphics[width=.47\textwidth]{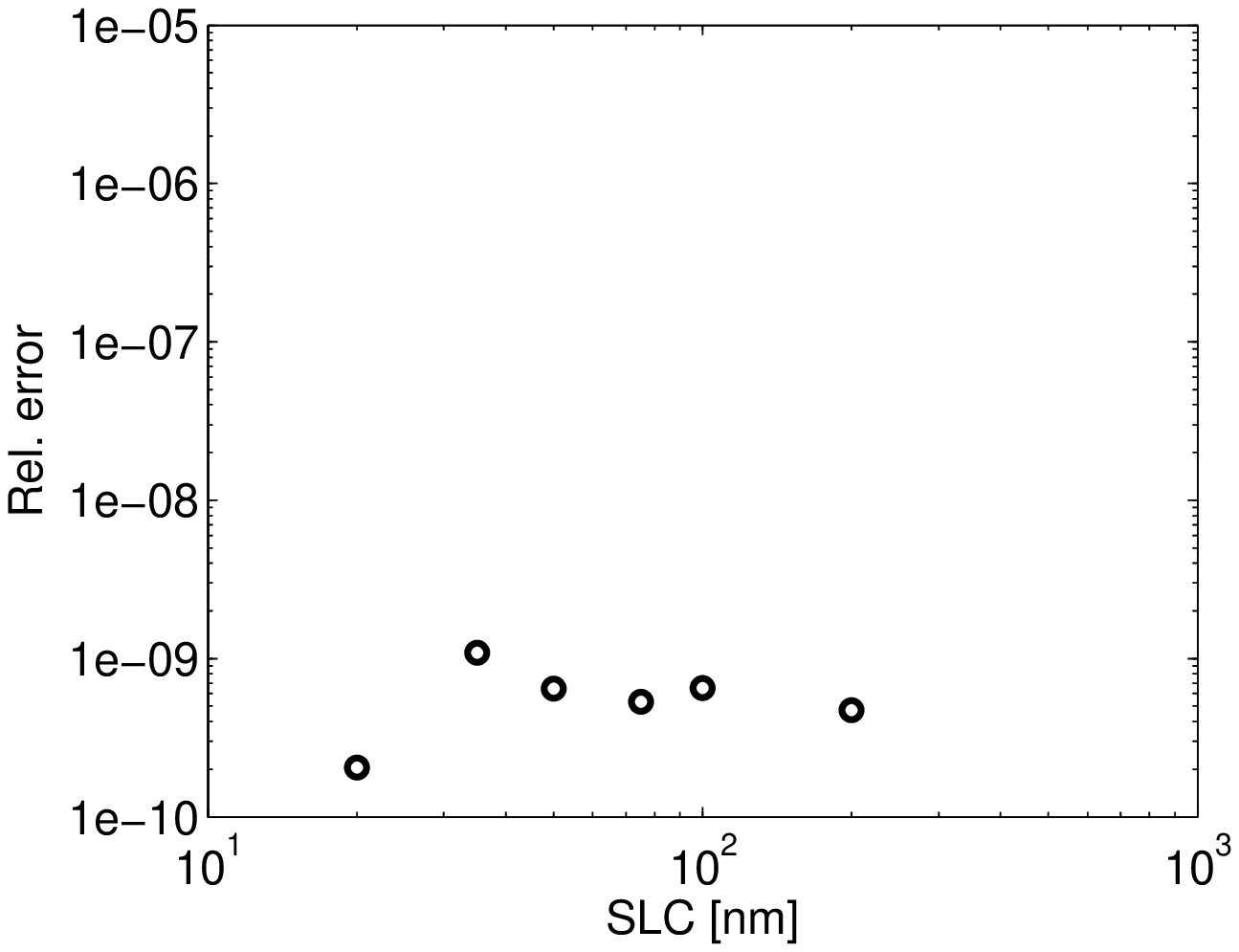}
  \hfill
  \includegraphics[width=.47\textwidth]{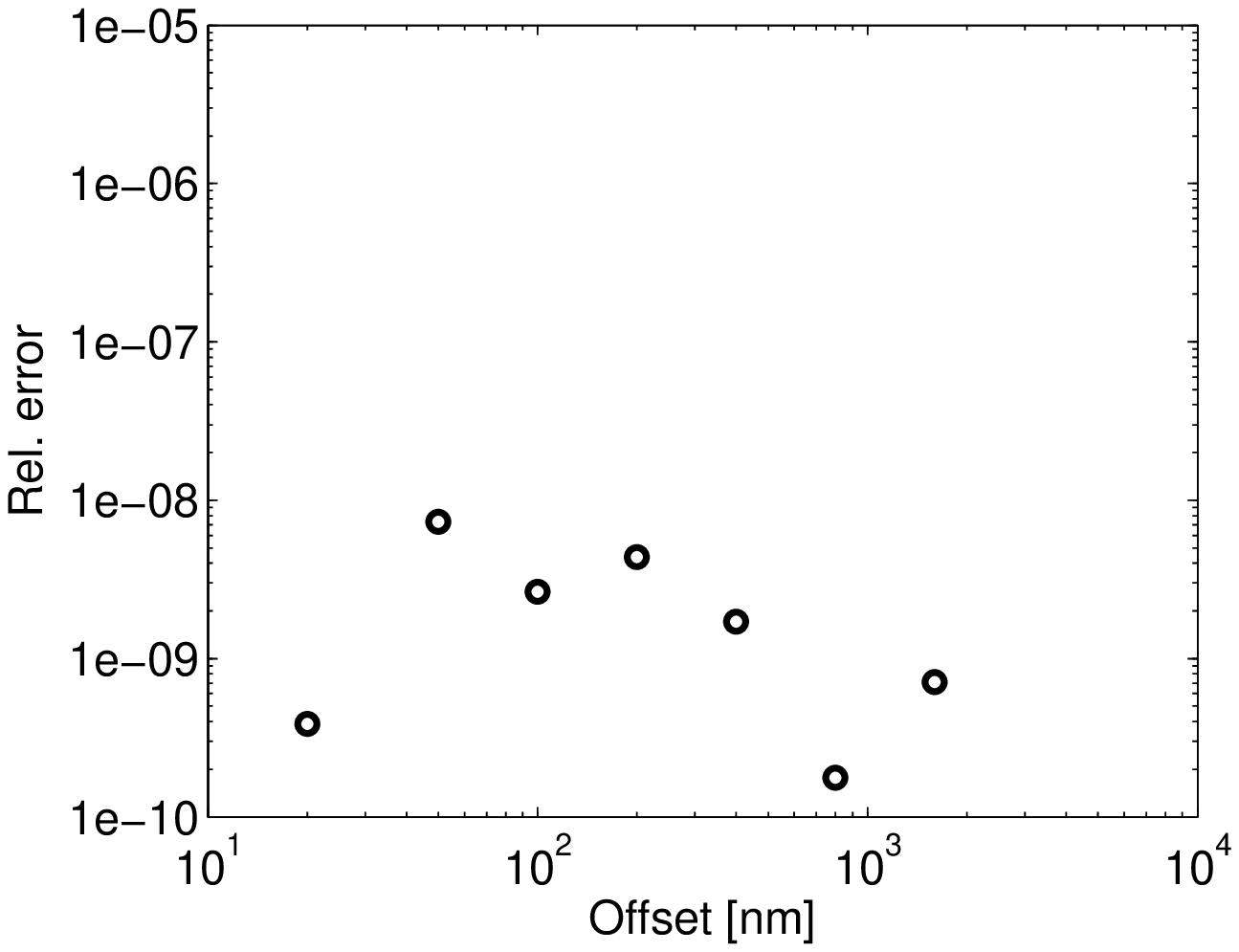}
  \caption{{\it Left:}  Relative numerical error, $\Delta(S/S_0)_{\mathrm{rel}}$, for various 
  settings of initial mesh sidelength constraints, $SLC$, and otherwise fixed numerical parameters.  
  {\it Right:}  Relative numerical error, $\Delta(S/S_0)_{\mathrm{rel}}$, for various 
  settings of offset of the computational domain boundary from the scattering structures, 
{\sffamily offset $w_\mathrm{x}$ }, and otherwise fixed numerical parameters.  As expected, 
the numerical error is dominated by the fixed numerical parameters. This shows that 
major contribution of $SLC$ and  offset $w_\mathrm{x}$ on the numerical error budget can 
be excluded.
  Please, compare also Table~\ref{table_my_results} (data sets V and VI).
}
\label{fig_convergence_slc_offs}
\end{center}
\end{figure}

{\bf Negligible influence of domain boundary placement and initial meshing} 
Finally Figure~\ref{fig_convergence_slc_offs} shows that neither the global initial mesh refinement, $SLC$
(Fig.~\ref{fig_convergence_slc_offs}, left), 
nor the distance of the computational domain boundary from the scattering structures, $dx$
(Fig.~\ref{fig_convergence_slc_offs}, right) has influence on the 
computational results larger than the error ranges which can be expected from 
used mesh refinement and finite element polynomial 
degree parameters, $n_\mathrm{adapt}+n_\mathrm{reg}$, $dx$, and $p$. 
The fact that the computational domain boundary placement has no effect on the numerical error
strongly indicates that influences of numerical parameters for realizing transparent 
boundary conditions are not present in this case.

{\bf Further tests} We have tested that using the Maxwell's time-harmonic wave equation for the electric or for the 
magnetic fields both give the same results within the numerical error ranges (which are similar for both methods). 
When the magnetic/electric field is computed directly, 
the electric/magnetic field necessary for flux-computation is obtained from 
numerical differentiation of the directly computed field. This
in principle could be subject to additional numerical errors. 
Therefore we have also performed simulation runs where we performed independent computations of 
both, electric and magnetic near fields and computed the electromagnetic flux through the 
detector from both of them. 
Also here we got the same quantitative results for $S/S_0$ and comparable magnitude of $\Delta(S/S_0)_\mathrm{rel}$. 
Also, rounding of the computational domain has no significant influence. 
Further, numerical parameters for the transparent boundary setting are obtained automatically and are different 
for each computation which makes influence of these parameters on the results beyond computed error estimates 
very improbable.

To summarize, we observe convergence of the results towards a value of $S/S_0 \approx 2.198825944
                                                                                  \pm 0.000000002$.
We observe convergence towards this value under increase of the polynomial degree of the used FEM 
ansatz functions. We also observe convergence towards the same value when the mesh is refined, where 
adaptive mesh refinement using a local error estimator, or (alternatively) using prerefinement of the
mesh around corners in the geometry is used. Also global mesh refinement shows convergence towards the 
same value. 
We observe neither influence of PML parameters nor influence of placement of the  computational 
domain boundary on the numerical results beyond those error limits which can be explained by the discretization 
parameters mesh refinement and FEM degree. 
Computation times for reaching relative accuracies of a level of $\Delta(S/S_0)_{\mathrm{rel}}\approx 10^{-6}$
are in the range of few seconds on a standard PC. 
Significantly better performance by orders of magnitude in comparison to the finite-element methods discussed 
in the literature~\cite{Lalanne2007jeos} can be explained (i) by the usage of higher-order edge-elements 
($p=2$ is used in~\cite{Lalanne2007jeos}), (ii) by usage of adaptive mesh refinement using automatic error-estimation 
(local refinement along metal-dielectric interfaces is used by FEM1 in~\cite{Lalanne2007jeos}, which is in contrast to 
error-estimation based automatic local mesh refinement which refines towards corners, 
regular meshes are used by FEM2 in~\cite{Lalanne2007jeos}), (iii) possibly by
the specific implementation of transparent boundary conditions.

\section{Conclusion}
Numerical convergence results from a Maxwell-solver based on a finite-element method 
have been presented for a challenging benchmark problem. 
The benchmark case treats rigorous light scattering in a non-periodic setup. 
Numerical performance and obtained accuracy are improved by several orders of magnitude when compared 
to results from the literature for this specific benchmark case. 
Fields of application of the method are, e.g.,  in scatterometric and ellipsometric 
metrology setups as well as in computational lithography. 
Highly efficient numerical performance is required in these fields for current and future 
technology node applications. 

\section*{Acknowledgments}
The work presented here is part of the EMRP Joint Research Project IND\,17 {\sc Scatterometry}.
The EMRP is jointly funded by the EMRP participating countries within EURAMET and the European Union.
We further acknowledge German research foundation, DFG, for funding within DFG research center {\sc Matheon}, 
project D23: {Design of nanophotonic devices and materials}.

\bibliography{/home/numerik/bzfburge/texte/biblios/phcbibli,/home/numerik/bzfburge/texte/biblios/my_group,/home/numerik/bzfburge/texte/biblios/lithography}

\bibliographystyle{spiebib}

\section*{Appendix}
\label{section_app}
The numerical results of this benchmark case using FEM solver JCMsuite are summarized in 
Table~\ref{table_my_results}. The table holds seven data sets: the numerical data setting for the 
quasi-exact result (see above), and for the parameter studies where each of the six numerical parameters 
has been varied independently. 
Results from data sets I to VI are displayed in Figures~\ref{fig_convergence_refinement}, 
\ref{fig_convergence_p_dx}, \ref{fig_convergence_slc_offs} .
The results from the literature 
are summarized in Table~\ref{table_results_lalanne}, where the relative errors have been estimated from the convergence graphs
given in the publication~\cite{Lalanne2007jeos}.

\begin{table}[h]
\begin{center}
\begin{tabular}{|l|l|l|l|}
\hline
method & & $S/S_0$ & est.~int.~error\\
\hline
MM1   & Aperiodic Fourier Modal Method (RCWA) & 2.206775 & $\pm$ 0.004\\
MM2   & Aperiodic Fourier Modal Method (RCWA) & 2.200904 & $\pm$ 0.0001\\
MM3   & Aperiodic Fourier Modal Method (RCWA) & 2.200952 & $\pm$ 0.000004\\
MM4   & Method of Lines (MOL) & 2.201970 & $\pm$ 0.0002\\
MM5   & Local Eigenmode-Modal method & 2.008785 & $\pm$ 0.002\\
FDTD1 & Finite-Difference Time-Domain Method & 2.211482 & $\pm$ 0.01\\
FDTD3 & Finite-Difference Time-Domain Method & 2.193380 & $\pm$ 0.04\\
FEM1  & Finite-Element Method & 2.201632 & $\pm$ 0.001\\
FEM2  & Finite-Element Method & 2.201143 & $\pm$ 0.000006\\
VIM   & Volume Integral Method & 2.204575 & $\pm$ 0.002\\
HYB   & Hybrid FEM-Modal Method & 2.200940 & $\pm$ 0.00002\\
\hline
\hline
FEM &  Finite-Element Method (this paper) &  2.198826 &  $\pm$ 1e-10 \\ 
\hline

\end{tabular}
\caption{
Results on $S/S_0$ and relative numerical error (with respect to best converged solution of each respective method) 
estimated from Fig.~5 in Lalanne {\it et al}~\cite{Lalanne2007jeos}.
For comparison, the last line of the table holds the quasi exact result from this paper. 
}
\label{table_results_lalanne}
\end{center}
\end{table}

\begin{table}[h]
\begin{center}
\begin{tabular}{|l|r|l|l||l|l|r|l|r|r|}
\hline
data set & $N$ & $S/S_0$ & $\Delta(S/S_0)_{\mathrm{rel}}$ & $p$ & $n_r$ & $n_a$ & $dx$\,[nm] & $SLC$\,[nm] & offset\,[nm]\\
\hline
Quasi-exact 
 &   1026290 & {\bf 2.1988259440} &  &  9 &  0 &  3 & 0.0010 & 50 & 200 \\ 
\hline
I 
 &    123282 & {\bf 2.1988}343722 & 3.83e-06 &  4 &  0 & \bf 0 & 0.2000 & 50 & 200 \\ 
 &    159612 & {\bf 2.19882}99667 & 1.83e-06 &  4 &  0 &  \bf 1 & 0.2000 & 50 & 200 \\ 
 &    174806 & {\bf 2.19882}78110 & 8.49e-07 &  4 &  0 &  \bf 2 & 0.2000 & 50 & 200 \\ 
 &    195658 & {\bf 2.19882}68085 & 3.93e-07 &  4 &  0 &  \bf 3 & 0.2000 & 50 & 200 \\ 
 &    221894 & {\bf 2.19882}63453 & 1.83e-07 &  4 &  0 &  \bf 4 & 0.2000 & 50 & 200 \\ 
 &    250338 & {\bf 2.19882}61217 & 8.08e-08 &  4 &  0 &  \bf 5 & 0.2000 & 50 & 200 \\ 
 &    286414 & {\bf 2.19882}60222 & 3.56e-08 &  4 &  0 &  \bf 6 & 0.2000 & 50 & 200 \\ 
 &    329320 & {\bf 2.1988259}599 & 7.27e-09 &  4 &  0 &  \bf 7 & 0.2000 & 50 & 200 \\ 
 &    384652 & {\bf 2.1988259}168 & 1.23e-08 &  4 &  0 &  \bf 8 & 0.2000 & 50 & 200 \\ 
 &    442726 & {\bf 2.1988259}068 & 1.69e-08 &  4 &  0 &  \bf 9 & 0.2000 & 50 & 200 \\ 
 &    527282 & {\bf 2.1988259}319 & 5.51e-09 &  4 &  0 & \bf 10 & 0.2000 & 50 & 200 \\ 
\hline
II 
 &    157184 & {\bf 2.198825}5506 & 1.79e-07 &  4 &  \bf 0 &  0 & 0.0100 & 50 & 200 \\ 
 &    622286 & {\bf 2.198825}6702 & 1.25e-07 &  4 &  \bf 1 &  0 & 0.0100 & 50 & 200 \\ 
 &   2308698 & {\bf 2.198825}8498 & 4.28e-08 &  4 &  \bf 2 &  0 & 0.0100 & 50 & 200 \\ 
 &   8489394 & {\bf 2.1988259}006 & 1.97e-08 &  4 &  \bf 3 &  0 & 0.0100 & 50 & 200 \\ 
 &  33140834 & {\bf 2.1988259}235 & 9.32e-09 &  4 &  \bf 4 &  0 & 0.0100 & 50 & 200 \\ 
 & 125886658 & {\bf 2.1988259}315 & 5.69e-09 &  4 &  \bf 5 &  0 & 0.0100 & 50 & 200 \\ 
\hline
III 
 &     82756 & {\bf 2.19}78707235 & 4.34e-04 &  \bf 1 &  0 &  3 & 0.0010 & 50 & 200 \\ 
 &    362394 & {\bf 2.1988}182593 & 3.49e-06 &  \bf 2 &  0 &  3 & 0.0010 & 50 & 200 \\ 
 &    783836 & {\bf 2.1988259}880 & 2.00e-08 &  \bf 3 &  0 &  3 & 0.0010 & 50 & 200 \\ 
 &    542996 & {\bf 2.1988259}107 & 1.51e-08 &  \bf 4 &  0 &  3 & 0.0010 & 50 & 200 \\ 
 &    432247 & {\bf 2.198825944}2 & 8.61e-11 &  \bf 5 &  0 &  3 & 0.0010 & 50 & 200 \\ 
 &    476798 & {\bf 2.198825944}2 & 1.13e-10 &  \bf 6 &  0 &  3 & 0.0010 & 50 & 200 \\ 
 &    637212 & {\bf 2.1988259440} & 2.43e-11 &  \bf 7 &  0 &  3 & 0.0010 & 50 & 200 \\ 
\hline
IV 
 &   8825980 & {\bf 2.19882}06540 & 2.41e-06 &  7 &  3 &  2 & \bf 5.0000 & 50 & 200 \\ 
 &  10121834 & {\bf 2.19882}70808 & 5.17e-07 &  7 &  3 &  2 & \bf 2.0000 & 50 & 200 \\ 
 &  11749026 & {\bf 2.19882}66893 & 3.39e-07 &  7 &  3 &  2 & \bf 1.0000 & 50 & 200 \\ 
 &  13364822 & {\bf 2.19882}60238 & 3.63e-08 &  7 &  3 &  2 & \bf 0.5000 & 50 & 200 \\ 
 &  17065246 & {\bf 2.19882}60053 & 2.79e-08 &  7 &  3 &  2 & \bf 0.2000 & 50 & 200 \\ 
 &  19191720 & {\bf 2.1988259}875 & 1.98e-08 &  7 &  3 &  2 & \bf 0.1000 & 50 & 200 \\ 
 &  21128326 & {\bf 2.1988259}844 & 1.84e-08 &  7 &  3 &  2 & \bf 0.0500 & 50 & 200 \\ 
 &  22332186 & {\bf 2.1988259}332 & 4.89e-09 &  7 &  3 &  2 & \bf 0.0200 & 50 & 200 \\ 
 &  23847490 & {\bf 2.19882594}10 & 1.35e-09 &  7 &  3 &  2 & \bf 0.0100 & 50 & 200 \\ 
 &  25328200 & {\bf 2.19882594}32 & 3.46e-10 &  7 &  3 &  2 & \bf 0.0050 & 50 & 200 \\ 
 &  28455268 & {\bf 2.198825944}8 & 3.95e-10 &  7 &  3 &  2 & \bf 0.0020 & 50 & 200 \\ 
 &  30191072 & {\bf 2.198825944}9 & 4.20e-10 &  7 &  3 &  2 & \bf 0.0010 & 50 & 200 \\ 
 &  30826532 & {\bf 2.198825944}3 & 1.72e-10 &  7 &  3 &  2 & \bf 0.0005 & 50 & 200 \\ 
\hline
V 
 &  24405586 & {\bf 2.19882594}50 & 4.73e-10 &  7 &  3 &  0 & 0.0010 & \bf 200 & 200 \\ 
 &  26543078 & {\bf 2.19882594}54 & 6.54e-10 &  7 &  3 &  0 & 0.0010 & \bf 100 & 200 \\ 
 &  28017866 & {\bf 2.19882594}51 & 5.32e-10 &  7 &  3 &  0 & 0.0010 & \bf 75 & 200 \\ 
 &  30172354 & {\bf 2.19882594}54 & 6.45e-10 &  7 &  3 &  0 & 0.0010 & \bf 50 & 200 \\ 
 &  36002178 & {\bf 2.19882594}63 & 1.08e-09 &  7 &  3 &  0 & 0.0010 & \bf 35 & 200 \\ 
 &  58449498 & {\bf 2.198825944}4 & 2.05e-10 &  7 &  3 &  0 & 0.0010 & \bf 20 & 200 \\ 
\hline
VI 
 &   8408962 & {\bf 2.198825944}8 & 3.87e-10 &  4 &  3 &  2 & 0.0100 & 50 & \bf 20 \\ 
 &   8278130 & {\bf 2.1988259}600 & 7.31e-09 &  4 &  3 &  2 & 0.0100 & 50 & \bf 50 \\ 
 &   7565714 & {\bf 2.19882594}98 & 2.64e-09 &  4 &  3 &  2 & 0.0100 & 50 & \bf 100 \\ 
 &   8499474 & {\bf 2.1988259}343 & 4.39e-09 &  4 &  3 &  2 & 0.0100 & 50 & \bf 200 \\ 
 &  10140418 & {\bf 2.19882594}02 & 1.71e-09 &  4 &  3 &  2 & 0.0100 & 50 & \bf 400 \\ 
 &  14758242 & {\bf 2.19882594}36 & 1.76e-10 &  4 &  3 &  2 & 0.0100 & 50 & \bf 800 \\ 
 &  27815378 & {\bf 2.19882594}24 & 7.08e-10 &  4 &  3 &  2 & 0.0100 & 50 & \bf 1600 \\ 
\hline
\end{tabular}
\caption{
Convergence study: FEM results for different numerical parameter settings. 
Number of unknowns, $N$, normalized energy flux, $S/S_0$, 
relative deviation from the {\it quasi-exact} result, $\Delta(S/S_0)_{\mathrm{rel}}$, 
FEM degree $p$, 
number of regular and adaptive mesh refinement steps, $n_r$ and  $n_a$, 
mesh prerefinement at corners, $dx$, mesh sidelength constraint,  $SLC$, 
computational domain boundary offset.
}
\label{table_my_results}
\end{center}
\end{table}

\end{document}